\documentclass[useAMS,usenatbib]{mn2e}
\usepackage{graphicx,subfigure}
\usepackage{pslatex}

\title[Stratification in blue horizontal-branch stars]{Stratification of the elements in the atmospheres of blue horizontal-branch stars}

\author[F. LeBlanc et al.]
{F. LeBlanc$^{1}$\thanks{E-mail: francis.leblanc@umoncton.ca}, A. Hui-Bon-Hoa$^{2}$, V.R. Khalack$^{1}$\\
$^{1}$D\'epartement de Physique et d'Astronomie, Universit\'e de Moncton, Moncton, N.-B., E1A 3E9, Canada\\
$^{2}$Laboratoire d'Astrophysique de Toulouse-Tarbes, Universit\'e de Toulouse, CNRS, 14 avenue Edouard Belin 31400, Toulouse, France}

\begin{document}

\date{Accepted 2010 July 22. Received 2010 June 14}

\pagerange{\pageref{firstpage}--\pageref{lastpage}} \pubyear{2010}

\maketitle

\label{firstpage}

\begin{abstract}{Blue horizontal-branch (BHB) stars with $T_{\rm eff}$ approximately larger than 11500 K show several observational anomalies. In globular clusters, they exhibit low rotational velocities, abundance anomalies (as compared to cluster abundances), photometric jumps and gaps and spectroscopic gravities lower than predicted by canonical models. It is commonly believed that the low rotational velocities of these stars permit atomic diffusion to be efficient in their atmosphere thereby causing the observed anomalies. Recent detections of vertical stratification of iron (and some other chemical elements) in several BHB stars concur with this framework. In this paper, improved model atmospheres that include the vertical stratification of the elements are applied to BHB stars to verify if they can explain their observational anomalies. The results from theoretical model atmospheres are consistent with the photometric jumps and gaps observed for BHB stars in globular clusters. It is found that iron stratification in the theoretical models and that obtained from observations have similar tendancies. Our results also show that the spectroscopic gravities obtained while using chemically homogeneous model atmospheres to fit observations are underestimated. These results significantly strengthen the belief that atomic diffusion is responsible for these BHB-star anomalies.}
\end{abstract}

\section{Introduction} 

Horizontal-branch stars are evolved intermediate-mass stars that are burning helium in their core (e.g. Hoyle \& Schwarzschild 1955). Some hot blue horizontal-branch (hereafter BHB) stars with $T_{\rm eff}$ larger than approximately 11500 K are very interesting objects since they exhibit several observational anomalies. This paper aims to study certain aspects of these astronomical objects in the light of recent modelling and observational results.

First, abundance anomalies are observed for such BHB stars in several globular clusters (Glaspey et al. 1989; Behr et al. 1999; Moehler et al. 1999; Behr et al. 2000a; Behr 2003a; Hubrig et al. 2009). Khalack et al. (2007, 2008 and 2010) have detected vertical stratification of the abundance of several chemical elements including Fe in the atmosphere of some stars of this type. Khalack et al. (2008) found that the iron abundance increases toward the lower atmosphere in three BHB stars. Nitrogen and sulfur stratification was also detected in the hot BHB star HD~135485 (Khalack et al. 2007). The abundances of these two elements are found to increase toward the upper atmosphere. More recently, Khalack et al. (2010) determined the vertical Fe abundance gradient in a sample of 14 BHB stars using the observed spectra of Behr (2003a). They detected vertical Fe stratification in five (and possibly seven) of these BHB stars. Isotopic anomalies have also been recently detected in BHB stars (Hubrig et al. 2009).

Photometric jumps are observed on the hot side of $T_{\rm eff} \simeq $ 11500 K for the horizontal-branch sequence of several globular clusters (Grundahl et al. 1999) as compared to what is predicted by canonical models. Photometric gaps are also observed for this sequence (Ferraro et al. 1998) at the same $T_{\rm eff}$ where the photometric jump mentioned above occurs.

Another anomaly detected is that the rotational velocities of BHB stars are observed to drop abruptly for $T_{\rm eff}$ approximately larger than 11500 K (Peterson, Rood \& Crocker 1995; Behr et al. 2000a and 2000b; Behr 2003b).

Finally, spectroscopic gravities of BHB stars in globular clusters, at least for the metal-poor ones (e.g. Crocker et al. 1988; Moehler et al. 1995), are also lower than those predicted by classical BHB models.

Atomic diffusion (Michaud 1970) is effective only if the medium is hydrodynamically stable enough to prevent mixing, because of the order of magnitude of the diffusion velocities which are much smaller than those of macroscopic motions. In the case of BHB stars, their relatively low rotational velocities should render their superficial layer stable enough for diffusion to take place. For instance, recently, Quievy et al. (2009) showed that for values of rotational velocities found for BHB stars with $T_{\rm eff}$ above 11500 K, meridional circulation in not efficient enough to prevent He from gravitationally settling. This leads to the disappearance of the superficial He convection zone and renders the atmosphere more stable. Once the medium is stable enough, atomic diffusion can dominate there, leading to abundance anomalies and vertical abundance stratification of the chemical elements through the migration of the particles caused by the disbalance of gravity and radiative forces, and yielding superficial abundance anomalies. Stratification of the elements has an effect on the physical structure of the atmosphere through the change in the opacities, thus leading to photometric anomalies, and to a change of the shape of the Balmer lines, explaining the low values of surface gravity obtained by fitting of these lines, if one uses canonical models. The scenario outlined above, where atomic diffusion takes place in the atmospheres of hot BHB stars, may therefore explain the various observational anomalies observed for these stars.
 
Hui-Bon-Hoa, LeBlanc and Hauschildt (2000) constructed stellar atmosphere models of BHB stars with vertical stratification of the elements, and self-consistent atmospheric structure. These models were successful in qualitatively reproducing the above-mentioned anomalies when assuming that atomic diffusion becomes efficient in BHB stars with $T_{\rm eff} > $ 11500 K (i.e. the temperature above which the vast majority of the BHB stars rotate slowly). These models have been improved recently (LeBlanc et al. 2009), and the aim of this paper is to apply these new models to BHB stars and to compare the results to observational data. Among the improvements, the convergence scheme to obtain self-consistent abundance stratification was ameliorated in these models. Also, the diffusion coefficients now used take into account the interaction between the ions and neutral hydrogen. Moreover, in these new models, upper and lower abundance limits were imposed to avoid that certain elements attain extremely large (for the test-particle approximation of the diffusion theory to still be valid) or small (to avoid numerical problems) abundances at certain depths (see Section 3 of LeBlanc et al. 2009 for more details).

First the new model atmospheres with elemental stratification will be briefly described. Theoretical vertical stratification of Fe in models of various $T_{\rm eff}$ will be compared to the observational results of Khalack et al. (2007, 2008 and 2010). The synthetic photometry obtained with these models will then be presented, checking if they are able to reproduce observed photometric jumps and gaps for globular cluster BHB sequences. Synthetic Balmer lines are also compared between models with different prescriptions, to show how the diffusion model can account for the low values of gravity derived when using canonical models.

\section{Model atmospheres with stratified abundances}

The process of elemental stratification relies on the diffusion velocity, which causes the migration of the different chemical elements within stars. The leading terms that cause this velocity are gravity and the radiative acceleration resulting from the momentum transfer between the radiation field and each chemical species (e.g. Gonzalez et al. 1995). This transfer depends on the opacity of the species under consideration and the local monochromatic radiation field, which, in turn, via the monochromatic opacities, depends on the local abundances of the different species. The stratification of the abundances is thus a time-dependent process and it is therefore complex to treat it thoroughly. As the radiative transfer equation has to be solved explicitly in stellar atmospheres, which are optically thin media, time-dependent calculations are not feasible at the present time. Studies of abundance stratification can however be performed assuming that an equilibrium state can be reached, when the diffusion velocity of each element is nil at each layer of the atmosphere. This is the framework within which our calculations are performed. Even though the elemental stratification profiles can possibly differ from those computed time-dependently, this approximation can be used to gauge the impact of vertical stratification of the abundances on the atmospheric structure and on certain observed quantities. For a more detailed discussion surrounding time-dependent diffusion and equilibrium solution, the reader is referred to Alecian \& Stift (2007) in which Ap stars are studied.

The model atomspheres presented here are calculated with a modified version of the PHOENIX code (Hauschildt, Allard \& Baron 1999) as described by LeBlanc et al. (2009). The models computed here are in LTE and include 39 elements (H-Ga, Kr-Nb, Ba and La).  To build the stratification profile for each element, the code seeks iteratively a solution where the atmospheric structure and the vertical abundance stratifications yield a nil diffusion velocity for each chemical species (i.e. the equilibrium solution). Indeed, as the abundance of each species changes locally, so do the opacities, and thus the atmospheric structure. The resulting model atmospheres are thus self-consistent from this point of view.

For the evaluation of the radiative accelerations, both bound-bound and bound-free transitions are considered for each species while assuming LTE (Hui-Bon-Hoa et al. 2002). As the radiative transfer is solved with the opacity sampling method, the radiative accelerations, along with the radiative flux, are calculated using a sufficiently fine frequency grid to ensure adequate precision (e.g. LeBlanc, Michaud \& Richer 2000). Several improvements were brought to the models presented here as compared to the models used in Hui-Bon-Hoa, LeBlanc \& Hauschildt (2000). For example, the convergence scheme used to compute the stratification of the abundances has been modified. Also, more precise diffusion coefficients are used. More details concerning the improvements brought to the atmospheric code are found in LeBlanc et al. (2009).

A series of BHB model atmospheres from $T_{\rm eff}$ = 11000 to 18000 K (see Tab.~\ref{tab:parameters}) were constructed while including abundance stratifications. The surface gravities and masses used here are those of the zero-age horizontal-branch (ZAHB) models of VandenBerg et al. (2000).

\begin{table}
\caption{Physical parameters of models.}
\label{tab:parameters}
\begin{tabular}{ l l l c c }\hline
\hline
$T_{\rm eff}$(K)        &  log $g$  &  $M (M_{\odot})$   \\\hline 
       
11000  &  3.78  &  0.663            \\
12000  &  3.95  &  0.643            \\
13000  &  4.11  &  0.625            \\
14000  &  4.26  &  0.609            \\ 
16000  &  4.52  &  0.581            \\
18000  &  4.74  &  0.561              \\\hline
\end{tabular}
\end{table}

\begin{figure}
\center
\vspace{31pt}
\includegraphics[scale=0.6]{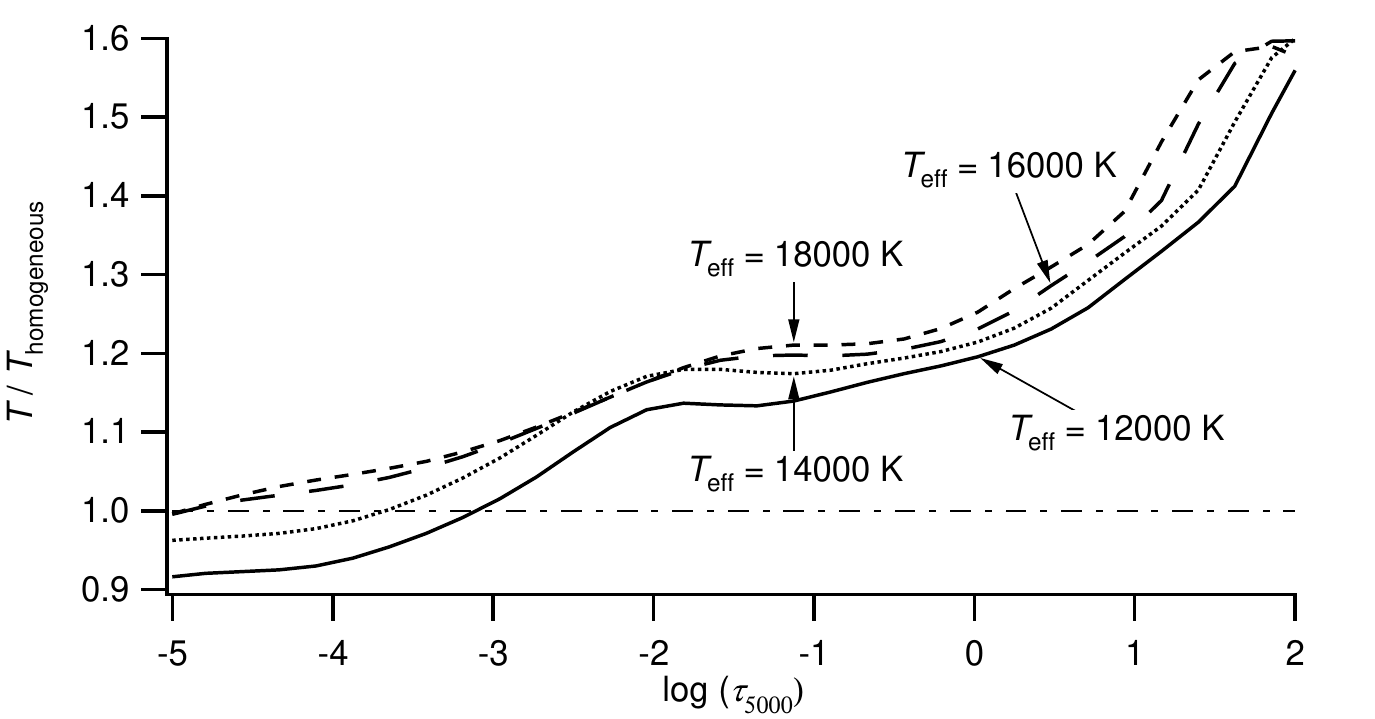}
\vspace{10pt}
\caption{The ratio of the temperature as a function of optical depth in various models including stratification to 
that of a chemically homogeneous model with metallicity of -1.5 dex solar of the same effective temperature and surface gravity. The dotted-dashed line shows the position of a ratio of one.}
\label{fig:Rapport_T}
\end{figure}

The stratification of the elements predicted at equilibrium strongly modifies the physical structure of the stellar atmospheres of hot BHB stars. Figure~\ref{fig:Rapport_T} shows the ratio of the temperature as a function of optical depth in several of these models as compared to the temperature of corresponding models (with the same fundamental parameters) with homogeneous models with a metallicity of -1.5 dex solar. The abundances used for these homogeneous models are chosen to be -1.5 dex compared to solar abundances which are typical of well-studied globular cluster metallicity (e.g. M13, M3 etc.). The large modifications in the structure of the self-consistent model atmospheres in the line-formation region as compared to homogeneous ones can give rise to detectable observational effects. As will be discussed below, the diffusion process not only causes abundance stratification of the elements but can also lead to visible effects related to the photometric colors and spectroscopic gravities of hot BHB stars. The results of these theoretical models are compared below to well-established anomalies of hot BHB stars. 

\section{Stratification of iron in BHB stars}

\begin{figure}
\center
\vspace{31pt}
\includegraphics[scale=0.6]{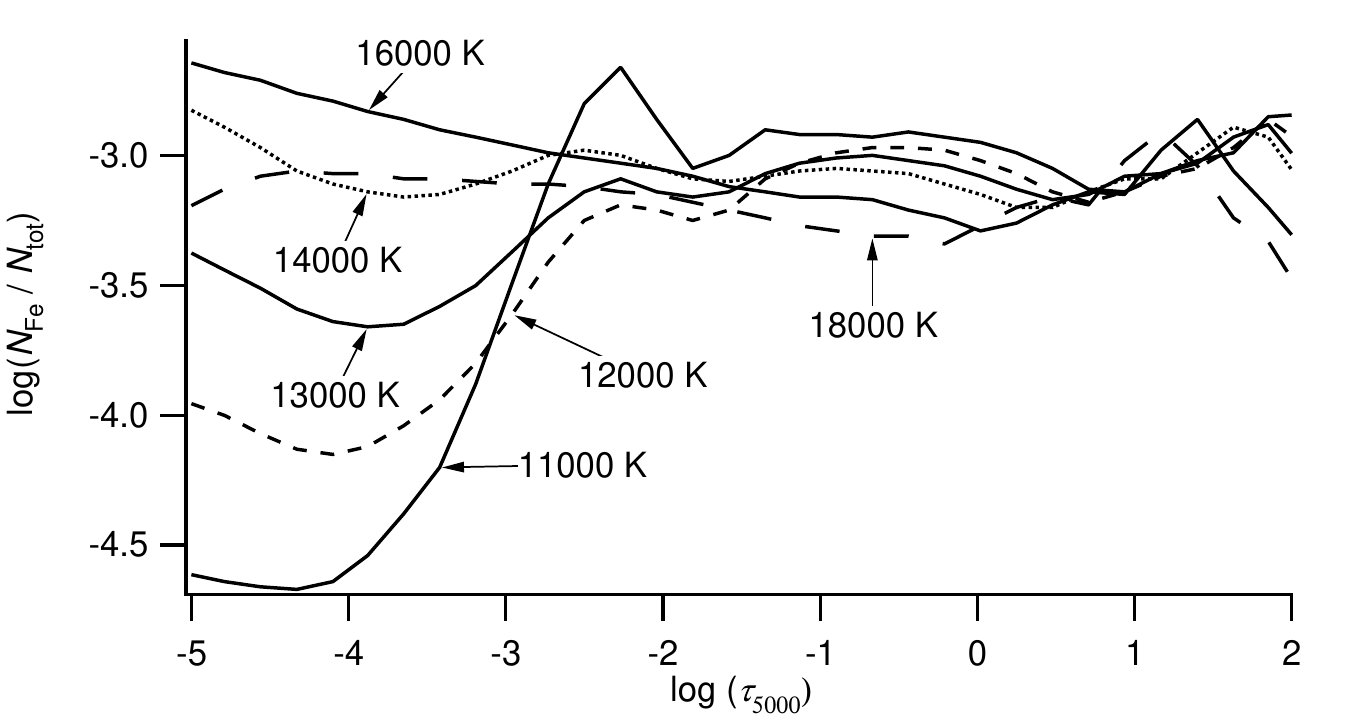}
\vspace{10pt}
\caption{Abundance of Fe relative to optical depth for self-consistent model atmospheres of BHB stars with $T_{\rm eff}$ from 11000 K to 18000 K.}
\label{fig:Fe_3dexmax_teff}
\end{figure}

The vertical stratification of Fe in our stratified model atmospheres is shown in Fig.~\ref{fig:Fe_3dexmax_teff}. The results presented in this figure clearly show that the gradient of the Fe abundance (as a function of optical depth in the atmosphere) in the range -4 $\preceq$ log~$\tau_{5000}$ $\preceq$ -2 decreases with the increase of $T_{\rm eff}$. Since this range of optical depth is where a lot of the iron lines are formed, such a tendancy can be verified with spectroscopic studies.

\begin{figure}
\center
\vspace{31pt}
\includegraphics[scale=0.6]{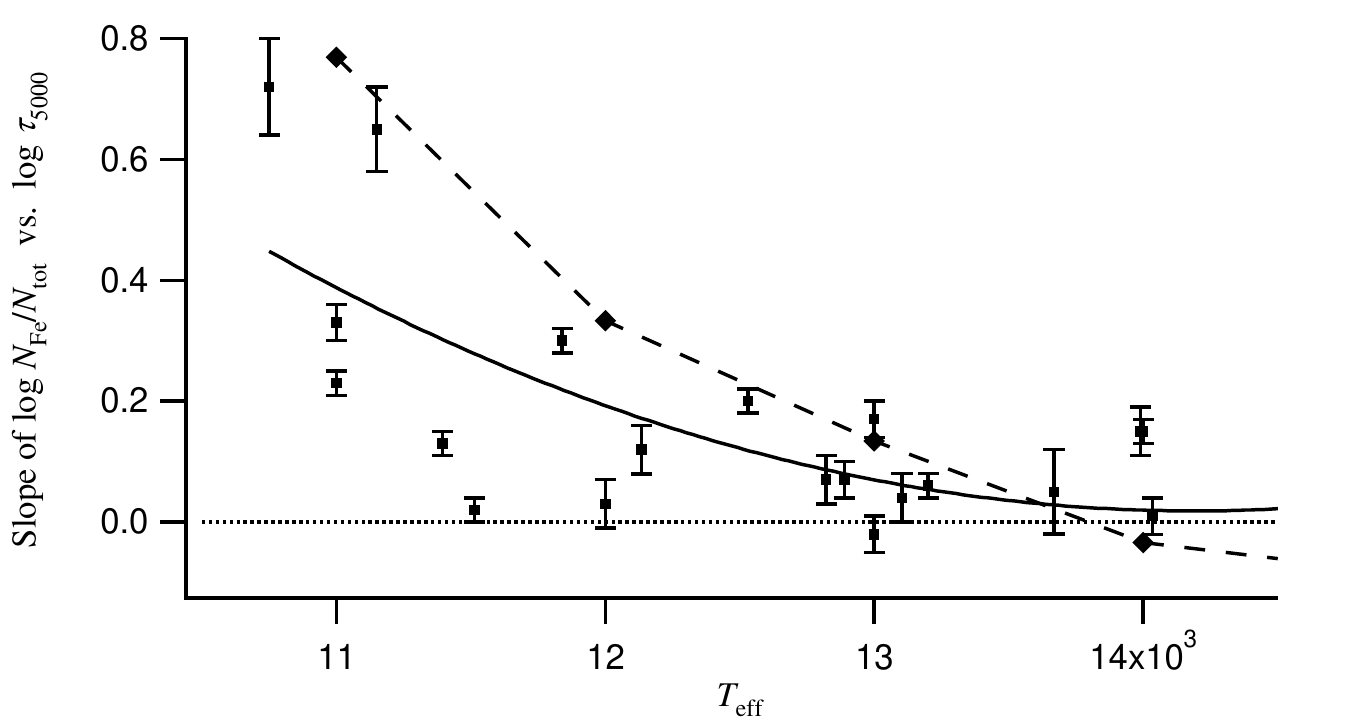}
\vspace{10pt}
\caption{Slope of the vertical Fe abundance obtained from the analysis of observed spectra (filled dots with their corresponding error bars) taken from Khalack et al. (2007, 2008 and 2010). Also shown in this figure is a quadratic fit of these observed slopes (solid line) and the linear slopes (filled diamonds joined by dashed line) obtained in the theoretical models in the range -5 $<$ log~$\tau_{5000}$ $<$ -2 which corresponds to the depth from where most of the Fe lines emanate. The dotted line shows the position of a nil slope.}
\label{fig:Slope_Fe}
\end{figure}

Figure~\ref{fig:Slope_Fe} compares the slopes of the detected Fe abundance gradients or lack thereof for a large number of BHB stars (Khalack et al. 2007, 2008 and 2010) to those predicted by the theoretical model atmospheres of LeBlanc et al. (2009). In these observational studies, the vertical stratification of iron was gauged by determining the abundance of individual lines formed at various optical depths. The dots in this figure represent individual hot BHB stars while the solid line is a quadratic fit of the observed abundance slopes. The filled diamonds connected by a dashed line give the linear slopes of Fe abundance in the theoretical models. These slopes are those of the linear fit of the abundance stratification of iron for the atmospheric layers in the range -5 $<$ log~$\tau_{5000}$ $<$ -2 for each $T_{\rm eff}$. These layers correspond to the atmospheric depths where most of the Fe lines are formed and where iron stratification is detected (see Figure 1 of Khalack et al. 2010). Figure~\ref{fig:Slope_Fe} shows that the observed slopes of the Fe abundance decreases with $T_{\rm eff}$ for hot BHB stars and eventually becomes undectectable at $T_{\rm eff}\simeq$ 14000 K, in agreement with the theoretical results of the models of LeBlanc et al. (2009) shown in Fig.~\ref{fig:Fe_3dexmax_teff}. These results may serve as additional proof that elemental stratification due to atomic diffusion occurs in the atmospheres of hot BHB stars.

\begin{figure}
\center
\vspace{31pt}
\includegraphics[scale=0.6]{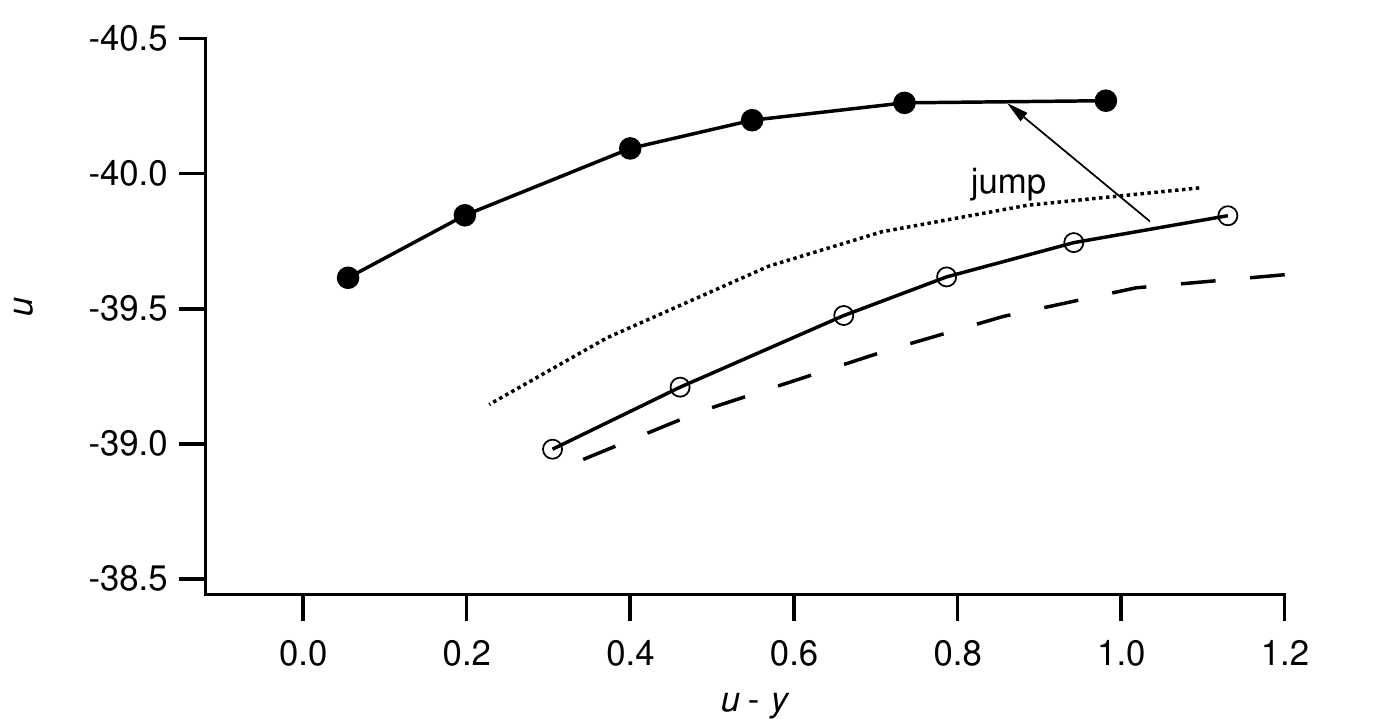}
\vspace{10pt}
\caption{Color-magnitude diagram ($u, u-y$) for the several sets of BHB-star models. The filled dots represent models from $T_{\rm eff} $ = 11000 to 18000 K described in Tab.~\ref{tab:parameters} ($T_{\rm eff}$ increases from right to left in this figure) that include vertical elemental stratification while the empty dots represent models with homogeneous abundances (with metals at -1.5 dex their solar values). An arrow shows the photometric jump predicted if diffusion becomes efficient at $T_{\rm eff} $ = 11500 K. The dotted curve represents atmosphere with homogeneous abundances equal to solar values. The dashed curve are the results when assuming stratification but without changing the structure of the atmosphere.}
\label{fig:jump}
\end{figure}

\begin{figure}
\center
\vspace{31pt}
\includegraphics[scale=0.6]{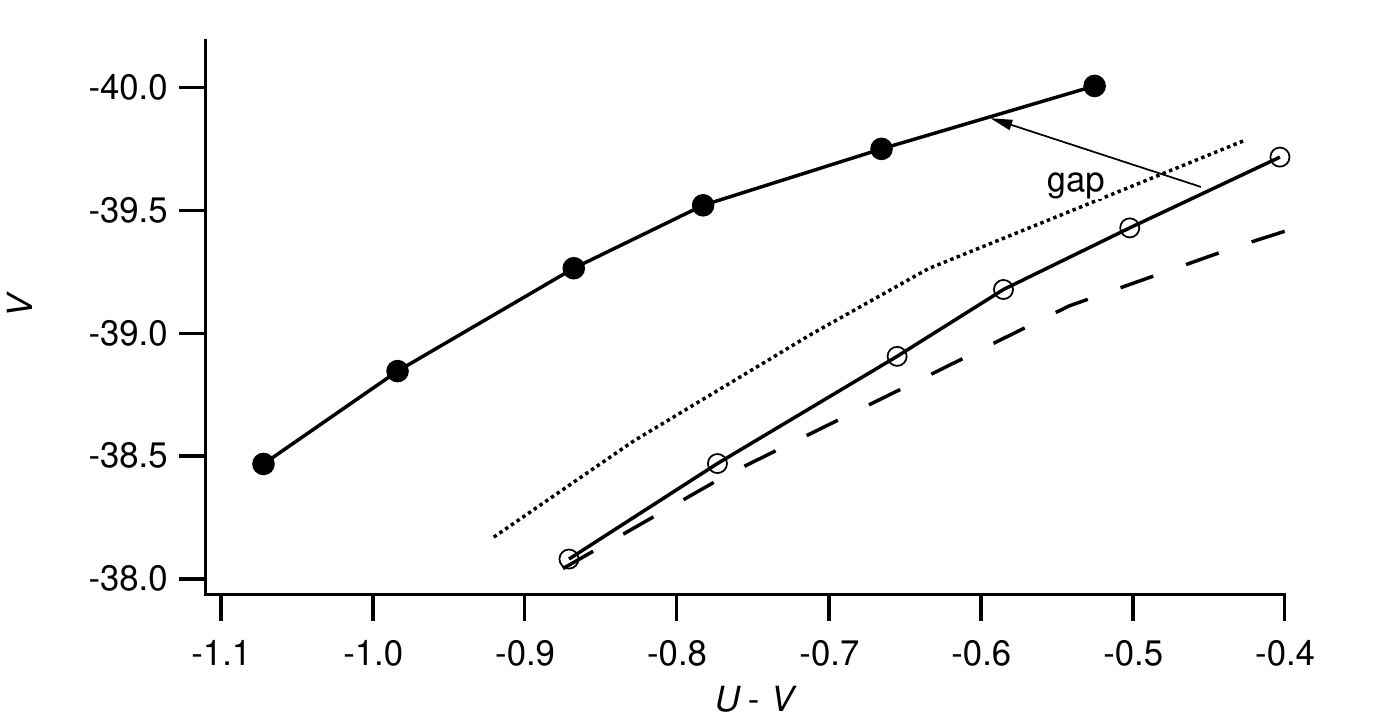}
\vspace{10pt}
\caption{Color-magnitude diagram ($V, U-V$) that shows the presence of a gap assuming that diffusion becomes efficient at $T_{\rm eff} $ = 11500 K. The curves are defined in the caption of Fig.~\ref{fig:jump}.}
\label{fig:gap}
\end{figure}

\section{Photometric jumps and gaps}

BHB stars in globular clusters exhibit several photometric anomalies namely photometric jumps and gaps. This section aims to interpret them in the light of self-consistent stratified model atmospheres.

Grundahl et al. (1999) observed a photometric jump in the ($u, u-y$) color-magnitude diagram for the horizontal branch sequence in several globular clusters. This jump consists of a shift towards brighter stars of the horizontal-branch sequence with respect to canonical models, and occurs for stars with $T_{\rm eff}$ higher than approximately 11500 K. Therefore, there exists a difference (for $T_{\rm eff} \succeq $ 11500 K) between the observed and predicted colors. The PHOENIX model atmospheres including elemental stratification of Hui-Bon-Hoa, LeBlanc \& Hauschildt (2000) were able to qualitatively explain this jump, when stars with $T_{\rm eff}$ larger than 11500 K become stable enough to allow atomic diffusion to have visible effects. This section aims to verify if the new models of LeBlanc et al. (2009) also reproduce this jump.

Figure~\ref{fig:jump} shows the ($u, u-y$) color-magnitude diagram for stars with and without elemental stratification. These results are consistent with the observational results of Grundahl et al. (1999) when compared to canonical models. Our results show that if diffusion becomes efficient at $T_{\rm eff} \simeq$ 11500 K, a jump occurs in the $(u,u-y)$ diagram. The difference in the {\it u}-bandpass magnitude between the homogeneous models (at -1.5 dex solar abundance for the metals) and those with stratification is 0.5 mag or more.  

Two other curves are also shown in this figure. The results while assuming that there is general enrichment of the metals (as compared to average cluster abundances) equal to a solar value is shown (dotted curve). This curve clearly shows that a general enrichment of the metals throughout the atmosphere cannot account for the photometric jump observed. The other curve shown in Figure~\ref{fig:jump} (dashed curve) represents the colors when one calculates the stratification profiles in the stars but without modifying the atmospheric structure (i.e. keeping the structure equal to that of the homogeneous model with -1.5 dex solar abundances). This shows that opacity changes cannot by themselves explain the photometric jump for BHB stars and that the structural changes are essential for this anomaly to occur. In summary, the photometric changes of the stratified models as compared to the homogeneous models are brought about by the change of the atmospheric structure due to the vertical stratification of the elements. For instance, the temperature in the line forming regions of hot BHB stars (see Figure~\ref{fig:Rapport_T}) found in the models with stratification is significantly larger (by up to approximately 20\%) as compared to the models with an homogeneous composition with -1.5 dex metal solar abundances.

Photometric gaps are also observed on the horizontal-branch sequence of globular clusters. These gaps in color-magnitude diagrams are characterized by an underpopulation of stars for specific values of colors.  For example, a photometric gap is observed in the ($V, U-V$) color-magnitude diagram (Ferraro et al. 1998). This gap corresponds to $T_{\rm eff} \simeq$ 11500 K where rotation plumets for BHB stars (Peterson, Rood \& Crocker 1995; Behr et al. 2000a and 2000b; Behr 2003b) and where diffusion could begin to be efficient. Hui-Bon-Hoa, LeBlanc and Hauschildt (2000) had shown that this gap may be explained if diffusion becomes efficient at $T_{\rm eff} \simeq$ 11500 K. The new models presented here also show the same feature (see Fig.~\ref{fig:gap}). The width of the gap is more than 0.1 mag which is consistent with the observations (see gap named G1 in Fig. 3 of Ferraro et al. 1998). As in Figure~\ref{fig:jump}, two other curves are plotted as well to show that the photometric gap cannot be explained by a homogeneous enrichment of the metals nor elemental stratification without structural changes.

\section{Spectroscopic gravities}
Several spectroscopic studies of BHB stars in metal-poor globular clusters give surface gravities lower than those predicted by canonical ZAHB models for stars with temperatures between 11,000 and 20,000~K (see Crocker et al. 1988; Moehler et al. 1995). The values of the surface gravities are determined through the study of Balmer lines profiles.

\begin{figure}
\begin{center}
\vspace{31pt}
\includegraphics[scale=0.45]{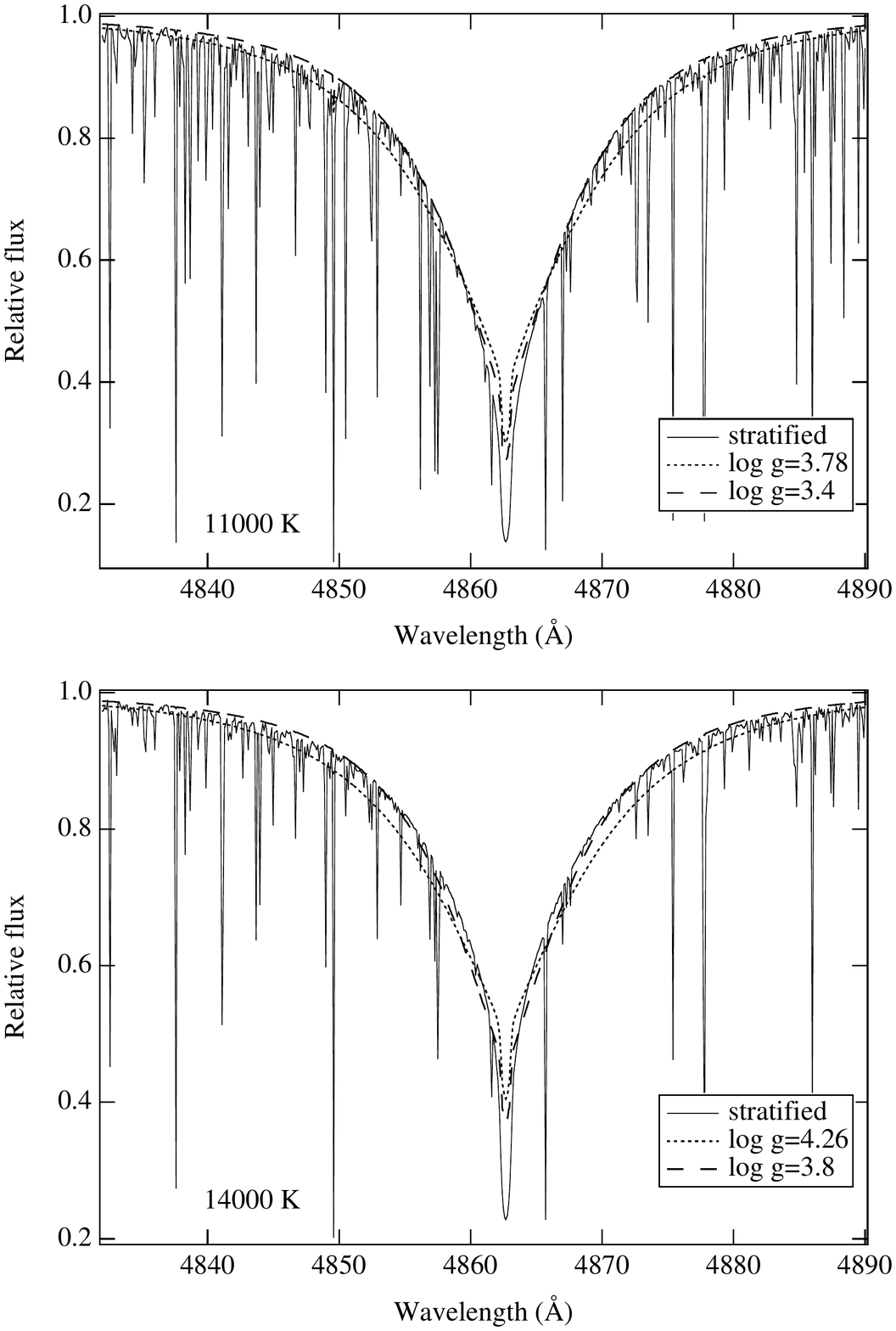}
\vspace{10pt}
\caption{$\mathrm{H}_\beta$ line profiles for the 11000 and 14000~K models. The stratified model profiles are denoted "stratified" and assume repectively $\log~g$ = 3.78 and 4.26 (see Tab.~\ref{tab:parameters}), whereas homogenous cluster abundance models (-1.5 dex solar) are named following their value of $\log~g$ used. The first value 
(dotted line) is that of the canonical sequence, and the second (dashed line) that of the homogenous model that fits best the stratified model.}
\label{fig:gravity}
\end{center}
\end{figure}

To see the effect of stratified model atmospheres on the value of surface gravities obtained this way, we synthesized Balmer lines for the self-consistent model atmospheres with the different effective temperatures used in the previous sections. Figure~\ref{fig:gravity} summarizes our results. Computed with values of surface gravity of canonical horizontal-branch models (see Tab.~\ref{tab:parameters}), the stratified models show Balmer line profiles closer to those of homogenous models of the same effective temperature but with lower gravity (by approximately 0.4 to 0.5~dex, see dashed line of Figure~\ref{fig:gravity}). For example, for the 14000 K model, the homogeneous model with $\log~g$ = 3.8 (dashed line) better fits the wings of the $\mathrm{H}_\beta$ line predicted by the self-consistent stratified model than the homogeneous model with the canonical value of $\log~g$ = 4.26 (dotted line). Thus, trying to determine spectroscopic gravities using homogenous models is misleading when the star has its structure altered by abundance stratification. This effect mimics lower spectroscopic gravities when using homogeneous models as a workbench. The use of model atmospheres taking into account abundance stratifications is therefore critical for spectroscopic gravity determinations.

As a test (not shown in Figure~\ref{fig:gravity}), we also computed Balmer line profiles using the atmospheric structure of the self-consistent stratified models, but with homogenous (cluster) abundances. The result is very close to that obtained with both the stratified model strucuture and abundances, showing that most of the effect arises from the change in the atmospheric structure induced by the stratification of the abundances.

\section{Discussion and conclusion}
The results presented in this paper show that model atmospheres calculated self-consistently with abundance stratifications (at equilibrium) of LeBlanc et al. (2009) successfully reproduce several observational anomalies of hot BHB stars. For instance, our theoretical models predict that the iron abundance stratification decreases as a function of $T_{\rm eff}$ and becomes negligible for BHB stars hotter than $T_{\rm eff}\simeq$ 14000 K. This result is consistent with the detected iron abundance slopes found by spectral line analysis of BHB stars undertaken by Khalack et al. (2007, 2008 and 2010).

The model atmospheres with vertical stratifications also reproduce the photometric jumps and gaps that occur for BHB stars at $T_{\rm eff}\simeq$ 11500 K. These anomalies were the precursors for the belief that abundance stratification due to diffusion is present in the slowly rotating hot BHB stars. The theoreticals models presented here also predict, as previously observed, that the spectroscopic gravities obtained for these stars while using homogeneous model atmospheres to fit observations systematically underestimate their true surface gravity. The photometric and surface gravity anomalies are due to the large modification in the atmospheric structure brought about by the stratification of the elements in the atmosphere.

In summary, the results presented here reaffirm those found by Hui-Bon-Hoa, LeBlanc \& Hauschildt (2000) regarding the photometric jumps and gaps as well as for the lower spectroscopic gravities found in hot BHB stars. In addition, we have found that the general trend for the observed Fe stratification slopes in hot BHB stars is similar to the one predicted by the theoretical models (i.e. the slope of Fe abundance decreases as $T_{\rm eff}$ increases).

The theoretical models used here assume that the elements attain an equilibrium where their respective diffusion velocities tend towards zero. The approach used to calculate the model atmospheres is not a time-dependent treatment of the diffusion phenomena. Such time-dependent modelling of the atmosphere is essential to obtain more precise elemental stratification profiles. Since Michaud et al. (2008) have calculated evolutionary models for BHB stars while including atomic diffusion, their results could be useful to constrain the amount of each element exchanged with the stellar interior, when building up time-dependent elemental stratifications in the atmospheres.

Even with the uncertainties inherent in the treatment of the diffusion phenomenon in the theoretical models used here, the results presented in this paper significantly strengthens the belief that atomic diffusion is responsible for the aformentioned anomalies observed for hot BHB stars. In the near future, we plan to investigate the error engendered by the use of homogeneous models for the photometric determination of fundamental parameters of BHB stars. Such a study is critical for ensuring the use of proper fundamental parameters for this type of star.

\section*{Acknowledgments}
We thank Luc LeBlanc who has modified some computer codes used here. We also thank the R\'eseau Qu\'eb\'ecois de Calcul de Haute Performance (RQCHP) for computational ressources. This research was partially supported by Natural Sciences and Engineering Research Council of Canada (NSERC) and Facult\'e des \'Etudes Sup\'erieures et de la Recherche de l'Universit\'e de Moncton.

\bsp

\label{lastpage}

\end{document}